\documentclass[final,5p,times,twocolumn]{elsarticle}

\usepackage{amssymb}
\usepackage{amsmath}
\usepackage{booktabs}
\usepackage{multirow}
\usepackage[T1]{fontenc}
\usepackage[colorlinks=true,linkcolor=blue,citecolor=blue,urlcolor=blue]{hyperref}
\usepackage{orcidlink}
\usepackage[switch,pagewise]{lineno}
\usepackage{dblfloatfix}
\usepackage{placeins}
\graphicspath{{./}}

\biboptions{sort&compress}

\begin{document}


\begin{frontmatter}

\title{Interaction fingerprints in temperature-fluctuation cumulants from the QCD crossover to nuclear liquid--gas criticality}

\author[inst1]{Debasish Mallick\,\orcidlink{0000-0002-5414-2421}}
\ead{dmallick89302010@gmail.com}
\affiliation[inst1]{organization={Key Laboratory of Quark \& Lepton Physics (MOE) and Institute of Particle Physics, Central China Normal University},
            city={Wuhan},
            postcode={430079},
            country={China}}

\begin{abstract}
Temperature fluctuations of the matter produced in relativistic heavy-ion
collisions are related to event-by-event fluctuations of the mean transverse
momentum and, through the specific heat, to the QCD equation of state. We
calculate the temperature-fluctuation cumulants $c_2$, $c_3$, and $c_4$ in the
ideal hadron resonance gas (HRG), in an excluded-volume HRG consistent with
lattice QCD constraints on baryon repulsion, and in a van der Waals HRG that
reproduces the nuclear ground state. At zero baryon density all three models
agree with lattice QCD thermodynamics up to the chiral crossover and separate
above it. Along the chemical freeze-out curve the models remain close for
$\sqrt{s_{NN}}\gtrsim20$~GeV and separate strongly at lower energies, where
baryonic interactions dominate the thermal response. In the van der Waals
model the variance $c_2$ develops a minimum along the Widom line of the
nuclear liquid--gas transition and $c_3$ changes sign across it. Temperature
cumulants thus connect mean-transverse-momentum fluctuations to thermodynamic
structures in two regions of the QCD phase diagram.
\end{abstract}


\end{frontmatter}
\raggedbottom

\section{Introduction}
\label{s:intro}

Relativistic heavy-ion collisions produce strongly interacting matter over a
wide range of temperature and baryon density. Lattice QCD calculations show
that at vanishing baryon chemical potential the transition from hadronic
matter to the quark--gluon plasma is a smooth crossover. At large baryon
density, effective and functional approaches predict a first-order phase
transition that ends at a critical end point (CEP)
\cite{Stephanov:1998dy,Stephanov:1999zu,Bzdak:2019pkr,Pandav:2022vpx,Fu:2019hdw}. The RHIC beam-energy scan program
searches for this critical point through event-by-event fluctuations of
conserved charges, in particular high-order cumulants of net-proton
distributions \cite{STAR:2020tga,STAR:2021iop,STAR:2022vlo,STAR:2025zdq}. At
still larger baryon density and lower temperature, ordinary nuclear matter
provides a second critical region: the liquid--gas phase transition.
Distinguishing the thermodynamic structures of these two regions is an
important part of mapping the QCD phase diagram
\cite{Braun-Munzinger:2020jbk}.

Temperature fluctuations offer a complementary probe. For a system in contact
with a heat bath, the variance of the temperature is fixed by the specific
heat, $\langle(\Delta T)^2\rangle=T^2/C_V$
\cite{Landau:1980stat,Stodolsky:1995ds}. In heavy-ion collisions, temperature
fluctuations are accessed through event-by-event fluctuations of the mean
transverse momentum $\langle p_T\rangle$
\cite{Stodolsky:1995ds,Gavin:2003cb,Korus:2001au}, which have been measured
over a wide range of collision systems and energies
\cite{ALICE:2024apz,ATLAS:2024jvf,STAR:2005rxx,PHENIX:2003ccl,
NA49:2008fys,CERES:2003sap}, and have been used to estimate the specific heat
of the produced matter \cite{Basu:2016ibk}. Very recently, the STAR
collaboration reported a non-monotonic collision-energy dependence of
two-particle $p_T$ correlations in central fixed-target Au+Au collisions at
$\sqrt{s_{NN}}=3.0$--$7.7$~GeV, with a significance of about $5\sigma$
\cite{STAR:2026ptc}. The measured fluctuations also
contain geometric and radial-flow contributions. Recent analyses have made
considerable progress in separating these sources from the thermal component
\cite{Gardim:2020sma,Zhang:2025mtm}.

Recently, Chen \emph{et al.} extended this idea to temperature cumulants of
arbitrary order \cite{Chen:2025jaq}. Starting from the state function
$W=\Omega+TS$, they obtained the cumulants from successive temperature
derivatives of the pressure and evaluated them with the functional
renormalization group (fRG). A subsequent study reported non-monotonic
temperature cumulants across the chiral and deconfinement transitions in the
PNJL model \cite{Liu:2025pnjl}. The second cumulant measures the specific
heat, while the higher orders measure how rapidly the thermal response
changes with temperature. Temperature cumulants therefore reorganize the
equation of state into observables with increasing sensitivity to crossover
and critical structures.

On the hadronic side of the crossover, the natural description is the hadron
resonance gas (HRG). The ideal HRG describes lattice thermodynamics below the
crossover, but it contains no interactions beyond resonance formation.
Repulsive interactions are commonly included through excluded-volume (EV)
corrections \cite{Rischke:1991ke}, and the allowed baryonic excluded volume
is constrained by lattice QCD susceptibilities
\cite{Vovchenko:2016rkn,Karthein:2021ucw}. Adding a van der Waals (VDW)
attraction connects the same hadronic description to the nuclear ground state
and predicts a liquid--gas critical point at
$(T_c,\mu_c)\simeq(19.7,908)$~MeV
\cite{Vovchenko:2015vxa,Vovchenko:2016rkn,Samanta:2017yhh}. Interactions of
this type are known to change conserved-charge fluctuations much more
strongly than the pressure itself \cite{Vovchenko:2016rkn,Karthein:2021ucw}.
Temperature cumulants, however, have so far been computed only in fRG and
PNJL approaches \cite{Chen:2025jaq,Liu:2025pnjl}; to the best of our knowledge, no such
calculation exists in any HRG model, ideal or interacting.

In this work we calculate the temperature-fluctuation cumulants $c_2$ through
$c_6$ in the ideal HRG (IHRG), a lattice-constrained baryon-only EVHRG, and a
nuclear-matter VDW-HRG. We compare the three models with lattice QCD at zero
baryon density, follow them along the chemical freeze-out curve, and examine
the nuclear liquid--gas region, where the VDW-HRG develops a Widom line. A
sector decomposition into mesons and baryons identifies the origin of the
model differences, and $c_5$ and $c_6$ probe the limits of the approach.
Together, these results provide a hadronic baseline that can help identify
critical structures in temperature fluctuations.

\section{Temperature-fluctuation cumulants}
\label{s:form}

The state function $W=\Omega+TS=U-\mu_BN_B$ has density $w=Ts-p$ and natural
variables $(S,V,\mu_B)$ \cite{Chen:2025jaq}. For
$\Delta T=T-\langle T\rangle$, the first nontrivial cumulants are
\begin{align}
\langle(\Delta T)^2\rangle_c &= \langle(\Delta T)^2\rangle ,\nonumber\\
\langle(\Delta T)^3\rangle_c &= \langle(\Delta T)^3\rangle ,
\label{e:cumdef}\\
\langle(\Delta T)^4\rangle_c &= \langle(\Delta T)^4\rangle
 -3\langle(\Delta T)^2\rangle^2 .\nonumber
\end{align}
They follow from entropy derivatives of $w$,
$\langle(\Delta T)^n\rangle_c=T^{4n-4}\partial^nw/\partial s^n$
\cite{Chen:2025jaq}. We introduce the dimensionless pressure derivatives
\begin{equation}
\chi_n(T,\mu_B)=T^{n-4}
\frac{\partial^np}{\partial T^n}\bigg|_{\mu_B,\mu_S,\mu_Q},
\label{e:chin}
\end{equation}
and scaled cumulants
\begin{equation}
c_n=\frac{\langle(\Delta T)^n\rangle_c}{T^n}.
\label{e:cndef}
\end{equation}
The second, third, and fourth orders are
\begin{equation}
c_2=\frac{1}{\chi_2},\qquad
c_3=-\frac{\chi_3}{\chi_2^3},\qquad
c_4=3\frac{\chi_3^2}{\chi_2^5}-\frac{\chi_4}{\chi_2^4}.
\label{e:cn24}
\end{equation}
For completeness, the next two orders are
\begin{align}
c_5={}&-15\frac{\chi_3^3}{\chi_2^7}
 +10\frac{\chi_3\chi_4}{\chi_2^6}-\frac{\chi_5}{\chi_2^5},
\label{e:c5}\\
c_6={}&105\frac{\chi_3^4}{\chi_2^9}
-105\frac{\chi_3^2\chi_4}{\chi_2^8}
+\frac{10\chi_4^2+15\chi_3\chi_5}{\chi_2^7}
-\frac{\chi_6}{\chi_2^6}.
\label{e:c6}
\end{align}

At second order, Eq.~(\ref{e:cn24}) reduces to the Einstein--Landau--Lifshitz
relation $\langle(\Delta T)^2\rangle=T^2/C_V$ \cite{Landau:1980stat}, with
$\chi_2=c_V/T^3$. The variance $c_2$ can therefore be obtained directly from
the lattice specific heat. The sign of $c_3$ tells whether the thermal
response is rising or falling with temperature, and $c_4$ measures its
curvature. Note that, since Eqs.~(\ref{e:cn24})--(\ref{e:c6}) are nonlinear in the
$\chi_n$, cumulants of individual particle sectors do not add up to the full
cumulant.

\section{Hadronic models and lattice input}
\label{s:models}

We use the PDG2020 particle list with a mass cutoff of $2.6$~GeV, excluding
charmed hadrons. With antiparticles the list contains 433 states, each
treated with quantum statistics. For a hadron $i$ with mass $m_i$, degeneracy
$d_i$, and chemical potential $\mu_i=B_i\mu_B+S_i\mu_S+Q_i\mu_Q$, the ideal
pressure is
\begin{equation}
p_i^{\rm id}=\pm\frac{d_iT}{2\pi^2}\int_0^\infty dk\,k^2
\ln\!\left[1\pm e^{-(\sqrt{k^2+m_i^2}-\mu_i)/T}\right],
\label{e:pid}
\end{equation}
where the upper and lower signs correspond to fermions and bosons. The IHRG
pressure is the sum over all species.

In the baryon-only EVHRG, mesons remain ideal, while baryons and antibaryons
form two separate repulsive sectors; meson--meson, meson--baryon, and
baryon--antibaryon EV terms are absent \cite{Vovchenko:2016rkn}. For either
baryonic sector,
\begin{equation}
p_{B(\bar B)}=\sum_{i\in B(\bar B)}p_i^{\rm id}
\bigl(T,\mu_i-bp_{B(\bar B)}\bigr).
\label{e:ev}
\end{equation}
We take $b=1.0$~fm$^3$ as the central value and show the interval
$b=0.4$--$1.0$~fm$^3$ where appropriate. This range is compatible with
lattice constraints from conserved-charge fluctuations
\cite{Vovchenko:2016rkn,Karthein:2021ucw}.

The VDW-HRG keeps the same sector structure and adds attraction. For a baryon
or antibaryon sector,
\begin{align}
p&=p^{\rm id}(T,\mu^*)-an^2,\nonumber\\
n&=\frac{n^{\rm id}(T,\mu^*)}{1+b\,n^{\rm id}(T,\mu^*)},\label{e:vdw}\\
\mu^*&=\mu-bp-abn^2+2an.\nonumber
\end{align}
We use $a=329$~MeV\,fm$^3$ and $b=3.42$~fm$^3$, fixed by the saturation
density and binding energy of the nuclear ground state
\cite{Vovchenko:2015vxa,Vovchenko:2016rkn}. The resulting equation of state
has a nuclear liquid--gas critical point near
$(T_c,\mu_c)=(19.7,908)$~MeV \cite{Vovchenko:2015vxa,Vovchenko:2016rkn}. The temperature derivatives entering the cumulants pick up the
interaction physics of the VDW-HRG.

For lattice QCD we use the analytic HotQCD pressure parametrization
\cite{HotQCD:2014kol} and the Wuppertal--Budapest (WB) trace-anomaly
parametrization \cite{Borsanyi:2013bia}; their spread indicates the
dependence on the analytic representation. Published thermodynamic points
with uncertainties are taken from the HotQCD tables \cite{HotQCD:2014kol} and
the recent WB equation of state \cite{Borsanyi:2025wb}. Only $c_2=1/\chi_2$
is compared directly with published lattice specific-heat points; the higher
cumulants shown as lattice curves are derivatives of the analytic
parametrizations.

The chemical freeze-out curve is taken from Andronic \emph{et al.}
\cite{Andronic:2017pug}:
\begin{equation}
\begin{split}
T_{\rm fo}(\sqrt{s_{NN}})&=
\frac{158.4}{1+\exp[2.60-\ln(\sqrt{s_{NN}})/0.45]}~{\rm MeV},\\
\mu_B(\sqrt{s_{NN}})&=
\frac{1307.5}{1+0.288\sqrt{s_{NN}}}~{\rm MeV},
\end{split}
\label{e:freezeout}
\end{equation}
with $\sqrt{s_{NN}}$ in GeV. At each point we impose strangeness neutrality
and the nuclear charge-to-baryon ratio, $n_S=0$ and $n_Q=0.4n_B$, and
determine $\mu_S$ and $\mu_Q$ separately in each model. The temperature
derivatives in Eq.~(\ref{e:chin}) are then evaluated at fixed
$(\mu_B,\mu_S,\mu_Q)$ at that point.

\begin{figure*}[!t]
\centering
\includegraphics[width=0.74\textwidth]{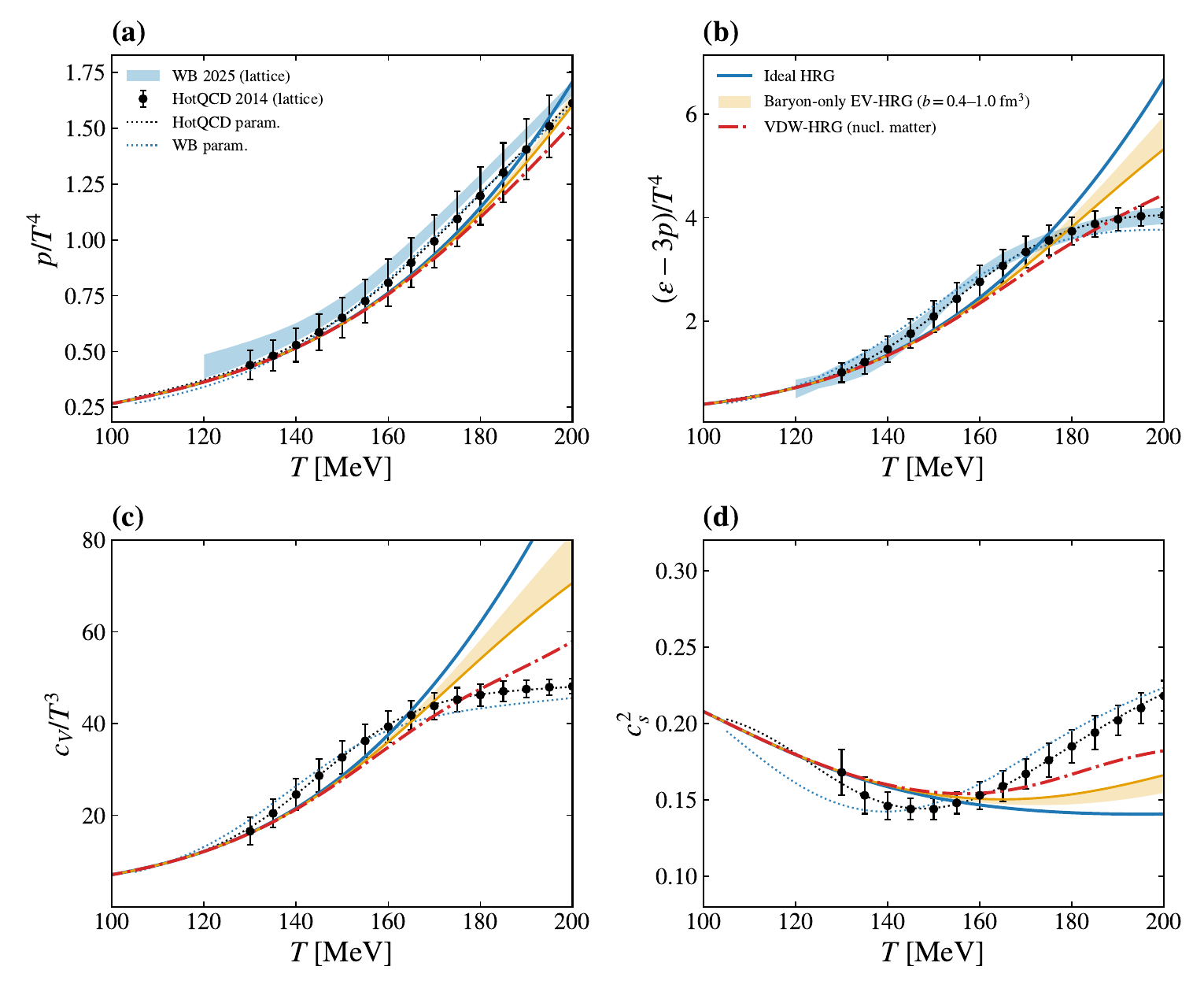}
\caption{Equation of state at $\mu_B=0$: (a) pressure, (b) trace anomaly,
(c) specific heat, and (d) speed of sound. Points with error bars are HotQCD data
\cite{HotQCD:2014kol}; the blue band is the WB data
\cite{Borsanyi:2025wb}; dotted curves are the HotQCD and WB analytic
parametrizations \cite{HotQCD:2014kol,Borsanyi:2013bia}. The model curves show
the IHRG, baryon-only EVHRG interval $b=0.4$--$1.0$~fm$^3$
\cite{Karthein:2021ucw}, and nuclear-matter VDW-HRG
\cite{Vovchenko:2016rkn}.}
\label{f:eos}
\end{figure*}

\section{Results and Discussion}
\label{s:res}

\subsection{Equation of state and lattice comparison}

Figure~\ref{f:eos} shows the pressure, trace anomaly, specific heat, and
speed of sound at $\mu_B=0$ for the three models, together with the lattice
data and parametrizations, in panels (a)--(d). The pressure in
Fig.~\ref{f:eos}(a) is nearly insensitive to the interaction prescription
below the crossover; its rapid rise is driven by the opening of hadronic
states. The derivative observables in panels (c) and (d) separate the models
much more clearly. The EV and VDW interactions soften the rise of $c_V/T^3$
relative to the IHRG and bend the speed of sound toward its minimum. This is
a first indication that temperature derivatives resolve interaction physics
that is hidden in $p/T^4$.

\begin{figure*}[!t]
\centering
\includegraphics[width=0.94\textwidth]{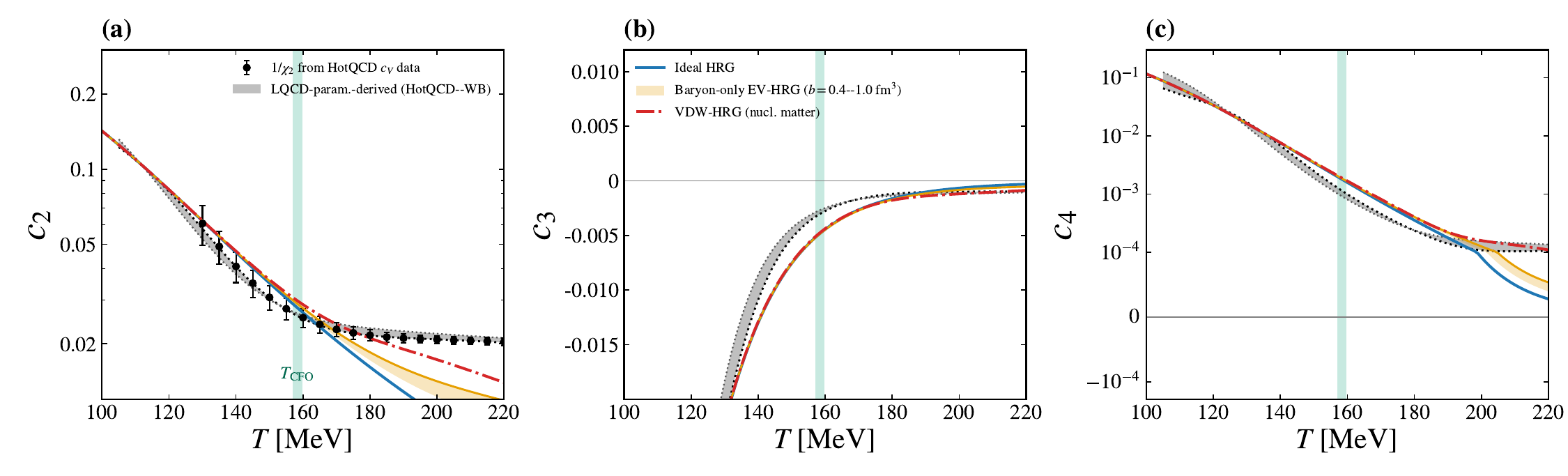}
\caption{Temperature-fluctuation cumulants (a) $c_2$, (b) $c_3$, and
(c) $c_4$ at $\mu_B=0$. Black points in panel (a) show $c_2=1/\chi_2$ from the published HotQCD
specific heat \cite{HotQCD:2014kol}, with propagated uncertainties. The grey
band spans results derived from the HotQCD and WB parametrizations
\cite{HotQCD:2014kol,Borsanyi:2013bia}. The HRG interactions follow
Refs.~\cite{Vovchenko:2016rkn,Karthein:2021ucw}. The vertical band marks
$T_{\rm CFO}=158.4\pm1.4$~MeV \cite{Andronic:2017pug}.}
\label{f:cnT}
\end{figure*}

The interacting models follow the lattice results up to
$T\simeq155$--$165$~MeV, including the development of the sound-speed
minimum. Above the crossover the fixed hadron spectrum cannot follow the
change in the active degrees of freedom, so the agreement is meaningful only
on the hadronic side. In particular, the flatter VDW-HRG specific heat at
high temperature reflects the suppression of the interacting baryon sector
while the mesons stay ideal; it is not a description of deconfined matter.

The reason interaction effects are weak in the pressure but visible in its
derivatives is the sector composition. Near $\mu_B=0$, ideal mesons carry
most of the thermal response, and the EV and VDW terms act only on the small
baryonic fraction. Each temperature derivative enhances the weight of the
rapidly changing baryon contribution. The sector-resolved calculation in
Appendix~\ref{a:sec1} shows this mechanism directly.

\subsection{Temperature cumulants across the crossover}

\begin{figure*}[!t]
\centering
\includegraphics[width=0.94\textwidth]{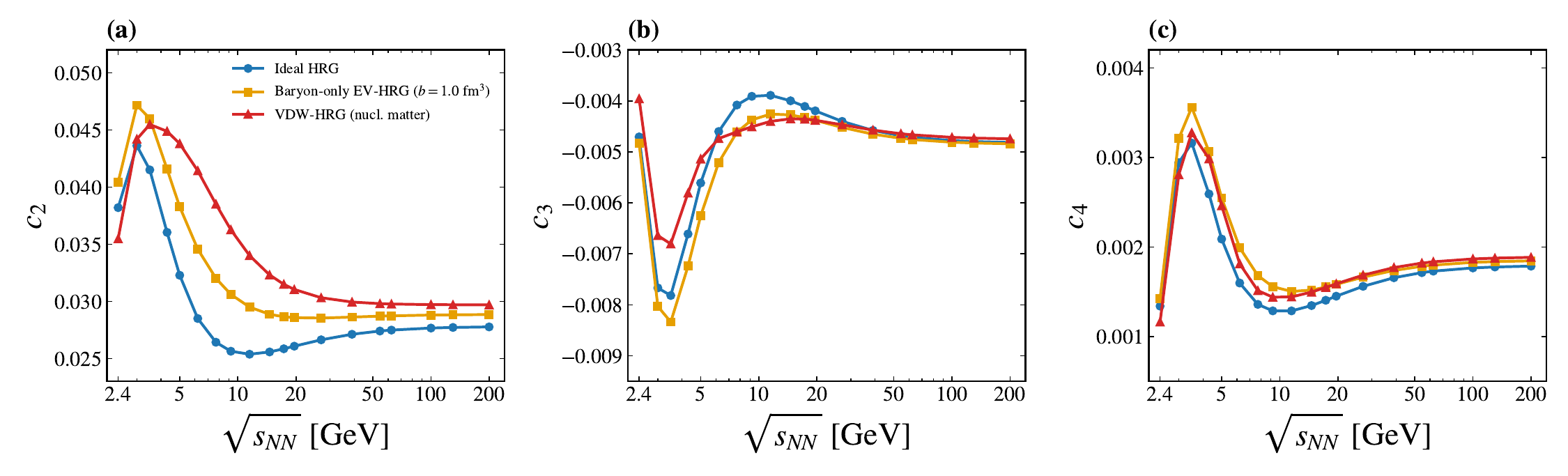}
\caption{Temperature-fluctuation cumulants (a) $c_2$, (b) $c_3$, and
(c) $c_4$ along the chemical freeze-out curve of Eq.~(\ref{e:freezeout}) \cite{Andronic:2017pug}, with $n_S=0$ and
$n_Q=0.4n_B$. The curves show the IHRG, baryon-only EVHRG with
$b=1.0$~fm$^3$, and nuclear-matter VDW-HRG
\cite{Vovchenko:2016rkn,Karthein:2021ucw}.}
\label{f:fo}
\end{figure*}

Figure~\ref{f:cnT} shows $c_2$, $c_3$, and $c_4$ at zero chemical potential
in panels (a)--(c). Below $T\simeq150$~MeV the three models and the lattice results are close,
because the gas is dilute and interaction corrections are small. At
$T=155$~MeV the models span $c_2=0.0305$--$0.0320$,
$c_3=-(5.70$--$5.80)\times10^{-3}$, and $c_4=(2.34$--$2.43)\times10^{-3}$,
consistent in scale with the fRG results of Ref.~\cite{Chen:2025jaq}. The
black points in Fig.~\ref{f:cnT}(a) are a direct conversion of the published
HotQCD specific heat and provide the cleanest lattice comparison in this
analysis.

The negative $c_3$ shows that the dimensionless specific heat rises through
this temperature range, and the positive $c_4$ describes the accompanying
curvature. Above the crossover the interactions bend the $c_2$ curves away
from the IHRG. We note that none of the HRG models produces a $c_2$
minimum, a $c_3$ sign change, or a $c_4$ upturn below 220~MeV. An
interaction-induced spread among hadronic models is therefore distinct from
the non-monotonic structures expected from a critical chiral mode.

The lattice comparison carries different weight at different orders. The
$c_2$ points inherit the published specific-heat uncertainties. The $c_3$ and
$c_4$ bands, in contrast, are obtained by differentiating the analytic
HotQCD and WB parametrizations, and their spread reflects the analytic
representation rather than a full lattice error budget. This distinction
becomes more important for $c_5$ and $c_6$, discussed in
Appendix~\ref{a:sec2}.

\subsection{Chemical freeze-out and increasing baryon density}

\begin{table*}[!t]
\centering
\caption{Representative values from Figs.~\ref{f:cnT} and \ref{f:fo}. The
freeze-out entries use the Andronic \textit{et al.} parametrization in
Eq.~(\ref{e:freezeout}) \cite{Andronic:2017pug}, with $n_S=0$ and
$n_Q=0.4n_B$; the 2.4 and 3.0~GeV rows are values on its low-energy
continuation. The model order in each group is IHRG (I), EVHRG with
$b=1.0$~fm$^3$ (E), and VDW-HRG (V). The third- and fourth-order cumulants
are multiplied by $10^3$.}
\label{t:summary}
\footnotesize
\setlength{\tabcolsep}{4.5pt}
\renewcommand{\arraystretch}{1.35}
\begin{tabular}{cc|ccc|ccc|ccc}
\hline\hline
Trajectory & \multicolumn{1}{c|}{Point} & \multicolumn{3}{c|}{$c_2$}
& \multicolumn{3}{c|}{$10^3c_3$} & \multicolumn{3}{c}{$10^3c_4$}\\
& & I & E & V & I & E & V & I & E & V\\
\hline
\multirow{3}{*}{\shortstack{$\mu_B=0$\\[2pt] ($T$ in MeV)}}
& 145 & 0.0403 & 0.0408 & 0.0410 & $-9.905$ & $-9.893$ & $-9.816$ & 5.124 & 5.176 & 5.207\\
& 155 & 0.0305 & 0.0314 & 0.0320 & $-5.783$ & $-5.796$ & $-5.701$ & 2.338 & 2.395 & 2.434\\
& 165 & 0.0234 & 0.0248 & 0.0261 & $-3.421$ & $-3.466$ & $-3.376$ & 1.080 & 1.137 & 1.171\\
\hline
\multirow{5}{*}{\shortstack{Freeze-out\\[2pt] ($\sqrt{s_{NN}}$ in GeV)}}
& 2.4 & 0.0382 & 0.0404 & 0.0355 & $-4.706$ & $-4.832$ & $-3.956$ & 1.340 & 1.423 & 1.164\\
& 3.0 & 0.0436 & 0.0472 & 0.0442 & $-7.667$ & $-8.030$ & $-6.641$ & 2.941 & 3.211 & 2.807\\
& 5.0 & 0.0323 & 0.0383 & 0.0438 & $-5.609$ & $-6.254$ & $-5.139$ & 2.089 & 2.548 & 2.462\\
& 7.7 & 0.0264 & 0.0320 & 0.0385 & $-4.080$ & $-4.615$ & $-4.606$ & 1.359 & 1.681 & 1.513\\
& 19.6 & 0.0261 & 0.0286 & 0.0310 & $-4.195$ & $-4.381$ & $-4.379$ & 1.451 & 1.583 & 1.589\\
\hline\hline
\end{tabular}
\end{table*}

\begin{figure*}[!t]
\centering
\includegraphics[width=0.94\textwidth]{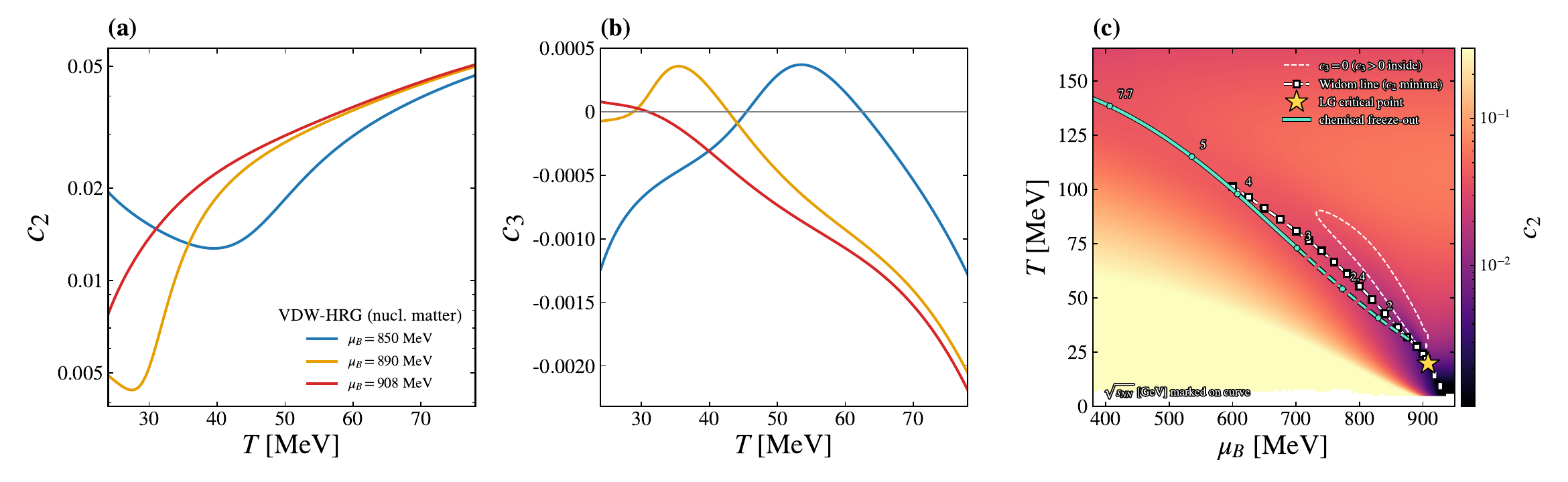}
\caption{Nuclear liquid--gas structure in the VDW-HRG
\cite{Vovchenko:2015vxa,Vovchenko:2016rkn}. Panels (a) and (b): $c_2$ and
$c_3$ versus temperature at $\mu_B=850$, 890, and 908~MeV. Panel (c): color
map of $c_2(\mu_B,T)$ on a logarithmic scale. The valley of $c_2$ (dark band) traces
the Widom line, shown by the $c_2$ minima (squares), and terminates at the
liquid--gas critical point (star). The dashed contour encloses the narrow
region where $c_3>0$: it opens near $\mu_B\simeq745$~MeV and runs along the
high-temperature side of the Widom line down to the critical-point region;
everywhere else in the plotted plane $c_3<0$. The chemical freeze-out curve of
Eq.~(\ref{e:freezeout}) \cite{Andronic:2017pug} is overlaid, with its
continuation below $\sqrt{s_{NN}}=3$~GeV dashed and collision energies in
GeV marked along the curve. The map and the Widom locus are evaluated at
$\mu_S=\mu_Q=0$; imposing $n_S=0$ and $n_Q=0.4n_B$ raises the
$c_2$-minimum temperatures by only $0.6$--$3.9$~MeV over
$\mu_B=700$--$905$~MeV and leaves the pattern unchanged. The low-temperature
first-order region, where $c_2$ is ill-defined along the metastable
branches, is masked in grey.}
\label{f:widom}
\end{figure*}

Figure~\ref{f:fo} shows the cumulants along the chemical freeze-out curve,
where temperature and chemical potentials change together; panels (a), (b),
and (c) show $c_2$, $c_3$, and $c_4$. From LHC energies down to
$\sqrt{s_{NN}}\simeq20$~GeV the three models remain close and nearly energy
independent, with $c_2\simeq0.026$--$0.031$ in Fig.~\ref{f:fo}(a),
$c_3\simeq-4\times10^{-3}$ in Fig.~\ref{f:fo}(b), and
$c_4\simeq1.5\times10^{-3}$ in Fig.~\ref{f:fo}(c). In this region mesons
carry most of the thermal response and the baryon density is too small for
the interactions to matter.

All three panels develop a pronounced structure at low collision energy. In
Fig.~\ref{f:fo}(a), $c_2$ rises below $\sqrt{s_{NN}}\simeq10$~GeV, reaches a
maximum near 3~GeV, and turns over on the low-energy continuation. The
higher cumulants show their extrema at nearly the same energy: $c_3$ in
Fig.~\ref{f:fo}(b) reaches its most negative value, about
$-8\times10^{-3}$ in the EVHRG, and $c_4$ in Fig.~\ref{f:fo}(c) its maximum
of about $3\times10^{-3}$. This common structure reflects two competing
trends along the trajectory: the falling freeze-out temperature reduces the
thermal response $\chi_2$ and thereby raises $c_2$, while below about 3~GeV
the rapidly growing baryon density enhances the response again.

The interaction hierarchy differs between panels and changes with energy. In
Fig.~\ref{f:fo}(a) the VDW-HRG gives the largest $c_2$ between roughly 5 and
20~GeV, whereas at 3~GeV and below the EVHRG lies highest and the VDW curve
drops back toward the ideal result. In Fig.~\ref{f:fo}(b) the VDW-HRG has
the least negative $c_3$ at the lowest energies, and in Fig.~\ref{f:fo}(c)
the EVHRG gives the largest $c_4$. Repulsion reduces the baryonic phase
space and changes the temperature derivatives of the baryon pressure;
attraction acts in the opposite direction at low density but grows in
importance as freeze-out approaches the nuclear liquid--gas region. Because
each $c_n$ combines several $\chi_n$ in a nonlinear way, this competition
reverses the model ordering between panels and between energies.
Representative values are collected in Table~\ref{t:summary}.

Two observations connect these results to the recent STAR measurement of
two-particle $p_T$ correlations, which shows a non-monotonic energy
dependence in central collisions at $\sqrt{s_{NN}}=3.0$--$7.7$~GeV
\cite{STAR:2026ptc}. First, the extrema of all three cumulants in
Fig.~\ref{f:fo} fall inside this fixed-target window. Second, even the
ideal HRG is mildly non-monotonic along the freeze-out curve: $c_2$ in
Fig.~\ref{f:fo}(a) has a shallow minimum near $\sqrt{s_{NN}}\simeq11$~GeV,
produced entirely by the trajectory, on which the freeze-out temperature
saturates at high energy while $\mu_B$ continues to grow toward low energy.
A smooth hadronic baseline can therefore generate mild non-monotonicity
without any critical structure. Quantifying such thermal baseline features
\cite{Braun-Munzinger:2020jbk} can support the interpretation of the STAR
measurement and help isolate a possible critical component in the observed
$p_T$ correlations.

The sector decomposition in Appendix~\ref{a:sec1} identifies the origin of this
behavior. The meson response varies slowly with collision energy, while the
baryon response grows toward low energy, becomes comparable to the mesonic
one, and carries essentially all of the interaction dependence. The model
hierarchy in Fig.~\ref{f:fo} is thus a direct fingerprint of baryonic
interactions in the thermal response of the fireball.

\subsection{Nuclear liquid--gas criticality}

Figure~\ref{f:widom} shows the temperature cumulants of the VDW-HRG in the
nuclear liquid--gas region. In Fig.~\ref{f:widom}(a), $c_2$ develops a
minimum that deepens and moves to lower temperature as $\mu_B$ approaches
its critical value: the minimum is $c_2\simeq0.013$ at $T=39.5$~MeV for
$\mu_B=850$~MeV, deepens to $c_2\simeq4.4\times10^{-3}$ at $T=27.5$~MeV for
$\mu_B=890$~MeV, and reaches the $10^{-3}$ level as $\mu_B$ approaches the
critical point. The minimum marks the enhancement of the specific heat by
the liquid--gas transition.

Figure~\ref{f:widom}(b) shows the accompanying sign structure of $c_3$.
Each curve crosses zero twice. The lower zero lies close to the $c_2$
minimum and separates the rising from the falling side of the liquid--gas
peak; the upper zero, near $T=62$~MeV for $\mu_B=850$~MeV and 43~MeV for
890~MeV, marks the handover from the liquid--gas enhancement to the
resonance-driven rise of the thermal response at higher temperature. The
zeros of $c_3$ locate the extrema of $\partial^2p/\partial T^2$, whereas the
$c_2$ minimum locates the maximum of $\chi_2$; the explicit powers of $T$ in
Eq.~(\ref{e:chin}) displace the two by about 6~MeV at $\mu_B=850$~MeV, and
the displacement shrinks to about 1~MeV at 890~MeV as the structure sharpens
toward the critical point.

Figure~\ref{f:widom}(c) combines these features in the $(\mu_B,T)$ plane.
The valley of $c_2$ traces the Widom line from
$(\mu_B,T)=(600,101.3)$~MeV to $(905,21.9)$~MeV and terminates at the model
critical point near $(908.9,19.5)$~MeV, consistent with the critical
point of Refs.~\cite{Vovchenko:2015vxa,Vovchenko:2016rkn}. The dashed $c_3=0$ contour encloses
a narrow band of positive $c_3$ that opens near $\mu_B\simeq745$~MeV and
follows the high-temperature side of the Widom line down to the
critical-point region; this is the sign structure of Fig.~\ref{f:widom}(b)
seen across the phase diagram. The freeze-out curve and its low-energy
continuation remain in the $c_3<0$ region. The low-energy continuation of the
chemical freeze-out curve runs nearly parallel to the Widom line and only
$5$--$9$~MeV below it in temperature between $\sqrt{s_{NN}}=3.0$ and
2.0~GeV. HADES, STAR fixed-target, and CBM energies therefore probe a domain
where temperature cumulants may be sensitive to nuclear liquid--gas
thermodynamics. The same panel also suggests that the location in the phase diagram and the
joint pattern of several cumulant orders provide complementary information
for separating liquid--gas structures from possible chiral critical
behavior.

The curves in Figs.~\ref{f:cnT}--\ref{f:widom} are equilibrium temperature
cumulants, not the full measured $\langle p_T\rangle$ distribution.
Initial-state geometry, radial-flow fluctuations, acceptance, and finite
multiplicity contribute as well. A quantitative comparison with data should
therefore propagate the equilibrium response through realistic collision
dynamics and test several cumulant orders together, particularly in the
low-energy fixed-target domain where both the interaction separation and the
liquid--gas response are largest
\cite{Gavin:2003cb,Gardim:2020sma,Zhang:2025mtm}.

\section{Summary and Outlook}
\label{s:summary}

To summarize, we have presented the first calculation of
temperature-fluctuation cumulants in ideal, excluded-volume, and van der
Waals hadron resonance gas models, covering the QCD crossover at $\mu_B=0$,
the chemical freeze-out curve, and the nuclear liquid--gas region. At zero
baryon density the interacting models follow lattice QCD thermodynamics on
the hadronic side of the crossover; $c_2$ is fixed by the lattice specific
heat, while $c_3$ and $c_4$ respond to its slope and curvature. The models
separate above the crossover, where the hadronic description loses
applicability.

Along chemical freeze-out, the three models agree for
$\sqrt{s_{NN}}\gtrsim20$~GeV and separate strongly below about 15~GeV. The
sector analysis shows that this separation is carried almost entirely by the
baryons: mesons dominate the thermal response at high energy, while the
growing baryon density exposes repulsive and attractive interactions at low
energy. In the nuclear-matter VDW-HRG, a minimum of $c_2$ and a sign change
of $c_3$ trace the liquid--gas Widom line, and the low-energy continuation of
the freeze-out curve runs toward it. Fixed-target measurements are therefore
a natural place to search for this thermodynamic pattern.

The fifth- and sixth-order cumulants illustrate both the increased resolving
power and the increased fragility of high temperature derivatives. Direct
lattice determinations of these derivatives, including their correlations,
would put the comparison on much firmer ground. On the experimental side, the
next step is to embed the equilibrium response in realistic collision
dynamics and connect it to the cumulants of the measured
$\langle p_T\rangle$ distribution after thermal, geometric, and radial-flow
components are separated \cite{Gavin:2003cb,Gardim:2020sma,Zhang:2025mtm}.
The non-monotonic $p_T$ correlations reported by STAR in the fixed-target
range \cite{STAR:2026ptc} lie in the region where the interaction
separation and the liquid--gas response found here are largest; the present
cumulants may provide a useful thermodynamic input for the interpretation
of such measurements. It
would also be interesting to repeat the calculation with alternative
repulsive schemes and with trajectories constrained by finite-density lattice
QCD, to establish which of the features found here are robust.

\section*{Acknowledgments}
This work is supported in part by the National Natural Science Foundation of China under Grant No. 12525509, No. 12447102, and the National Key Research and Development Program of China under contract No. 2022YFA1604900, and the Fundamental Research Funds for the Central Universities (XJ2026000302). I would also gratefully acknowledge Central China Normal University (CCNU), Wuhan, China for the hospitality and financial support as post-doctoral fellowship.

\FloatBarrier
\bibliographystyle{elsarticle-num}
\bibliography{temp_fluct_arxiv_DM}

\clearpage
\onecolumn

\appendix
\setcounter{figure}{0}
\setcounter{dbltopnumber}{2}
\renewcommand{\thefigure}{A.\arabic{figure}}
\renewcommand{\theHfigure}{A.\arabic{figure}}

\section{Particle-sector diagnostics and higher orders}
\label{a:sector}

\subsection{Meson and baryon sectors}
\label{a:sec1}

To identify the origin of the model hierarchy, we evaluate the cumulants from
sector-restricted pressures. The meson pressure contains all $B_i=0$ states;
the baryon pressure contains the baryon and antibaryon sectors with the
interaction of the corresponding model. At $\mu_B=0$ all sectors are
evaluated at $\mu_S=\mu_Q=0$. Along freeze-out, $\mu_S$ and $\mu_Q$ are first
determined from the full model, and the sector pressures are evaluated at the
same $(T,\mu_B,\mu_S,\mu_Q)$. Since Eqs.~(\ref{e:cn24})--(\ref{e:c6}) are
nonlinear, the sector cumulants do not sum to the full cumulant; they
diagnose the response of each restricted equation of state.

\begin{figure*}[!htb]
\centering
\includegraphics[width=0.96\textwidth]{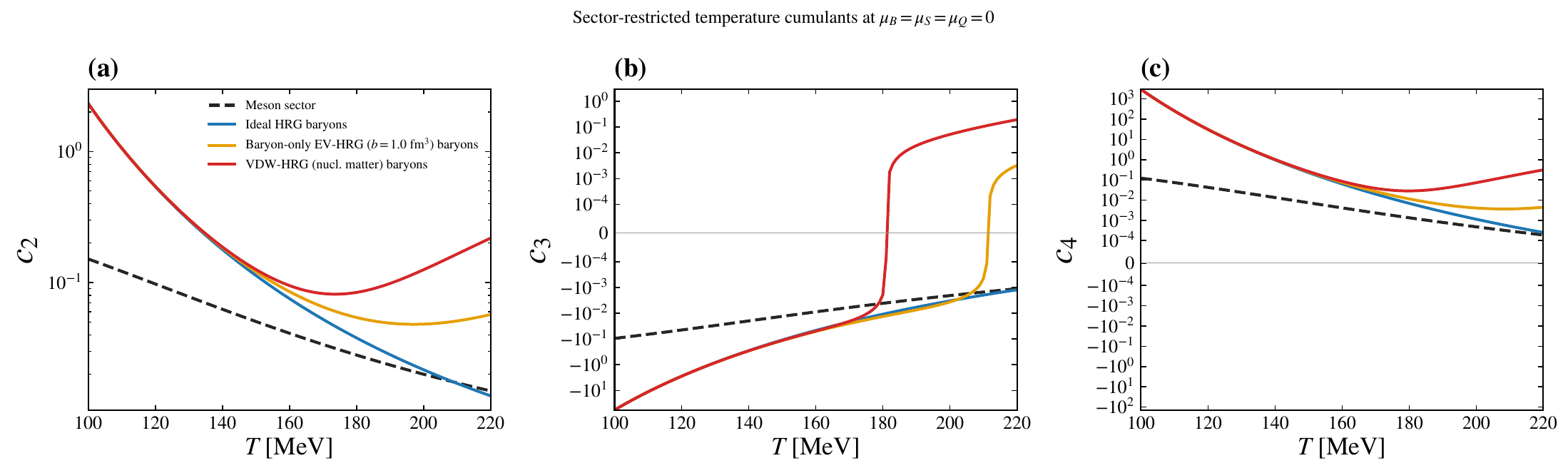}
\caption{Sector-restricted (a) $c_2$, (b) $c_3$, and (c) $c_4$ versus
temperature at $\mu_B=\mu_S=\mu_Q=0$. The common dashed curve is the ideal meson sector.
Solid curves are the baryon plus antibaryon sectors of the IHRG, baryon-only
EVHRG, and nuclear-matter VDW-HRG, using the interactions of
Refs.~\cite{Vovchenko:2016rkn,Karthein:2021ucw}. Sector cumulants are
diagnostic and are not additive components of the full $c_n$.}
\label{f:sectorT}
\end{figure*}

Figure~\ref{f:sectorT} shows the sector cumulants at zero chemical potential.
The meson curve is identical for all three models because mesons remain
ideal. Below about 150~MeV the baryon curves are also close, reflecting the
dilute gas. Their separation grows with temperature: EV repulsion bends the
baryon response away from the IHRG, and the VDW parameters change the higher
derivatives even more strongly. The large sector values at the edges of the
plotted range arise because the sector-specific $\chi_2$ becomes small there;
they are not additive enhancements of the full cumulant.

\begin{figure*}[!htb]
\centering
\includegraphics[width=0.96\textwidth]{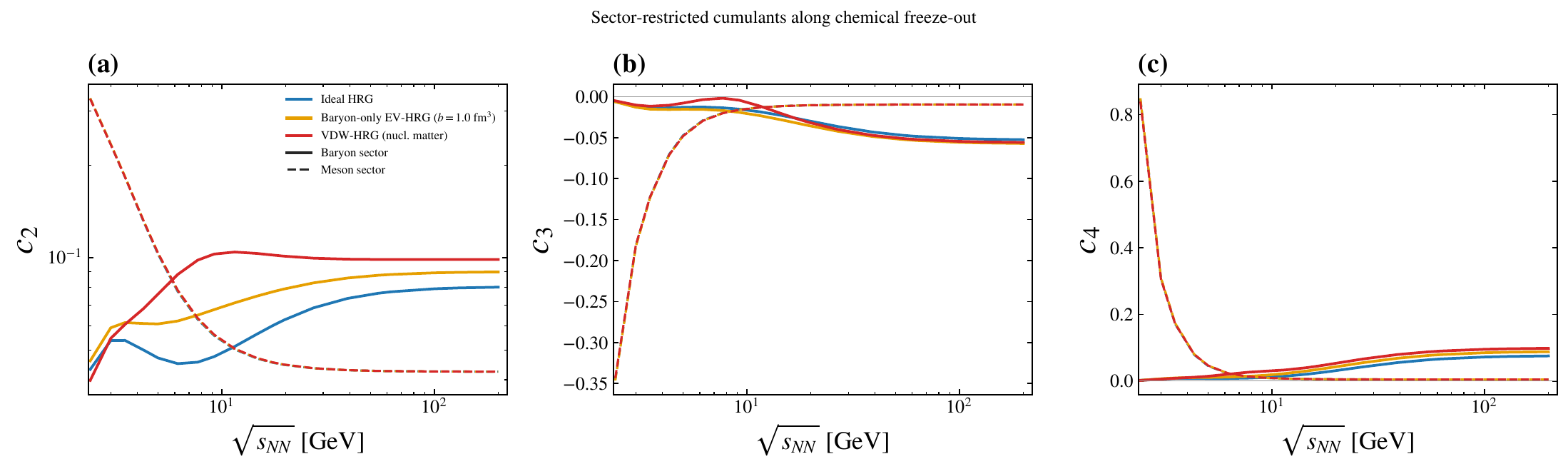}
\caption{Sector-restricted (a) $c_2$, (b) $c_3$, and (c) $c_4$ along the
chemical freeze-out curve of Eq.~(\ref{e:freezeout}) \cite{Andronic:2017pug}. Colors
identify the IHRG, baryon-only EVHRG, and nuclear-matter VDW-HRG
\cite{Vovchenko:2016rkn,Karthein:2021ucw}; solid and dashed curves denote the
baryon and meson sectors. Each sector is evaluated with the chemical
potentials obtained from the corresponding full model.}
\label{f:sectorFO}
\end{figure*}

Figure~\ref{f:sectorFO} shows the same decomposition along freeze-out. At
high collision energy the meson curves vary slowly, and the model differences
are controlled by a small baryonic component. Toward low energy the baryon
response becomes comparable to or larger than the mesonic one, and the ideal,
EV, and VDW baryon curves separate strongly. The full-model hierarchy in
Fig.~\ref{f:fo} follows from these changing sector weights combined with the
nonlinear structure of Eq.~(\ref{e:cn24}).

\subsection{Fifth- and sixth-order cumulants}
\label{a:sec2}

\begin{figure*}[!htb]
\centering
\includegraphics[width=0.82\textwidth]{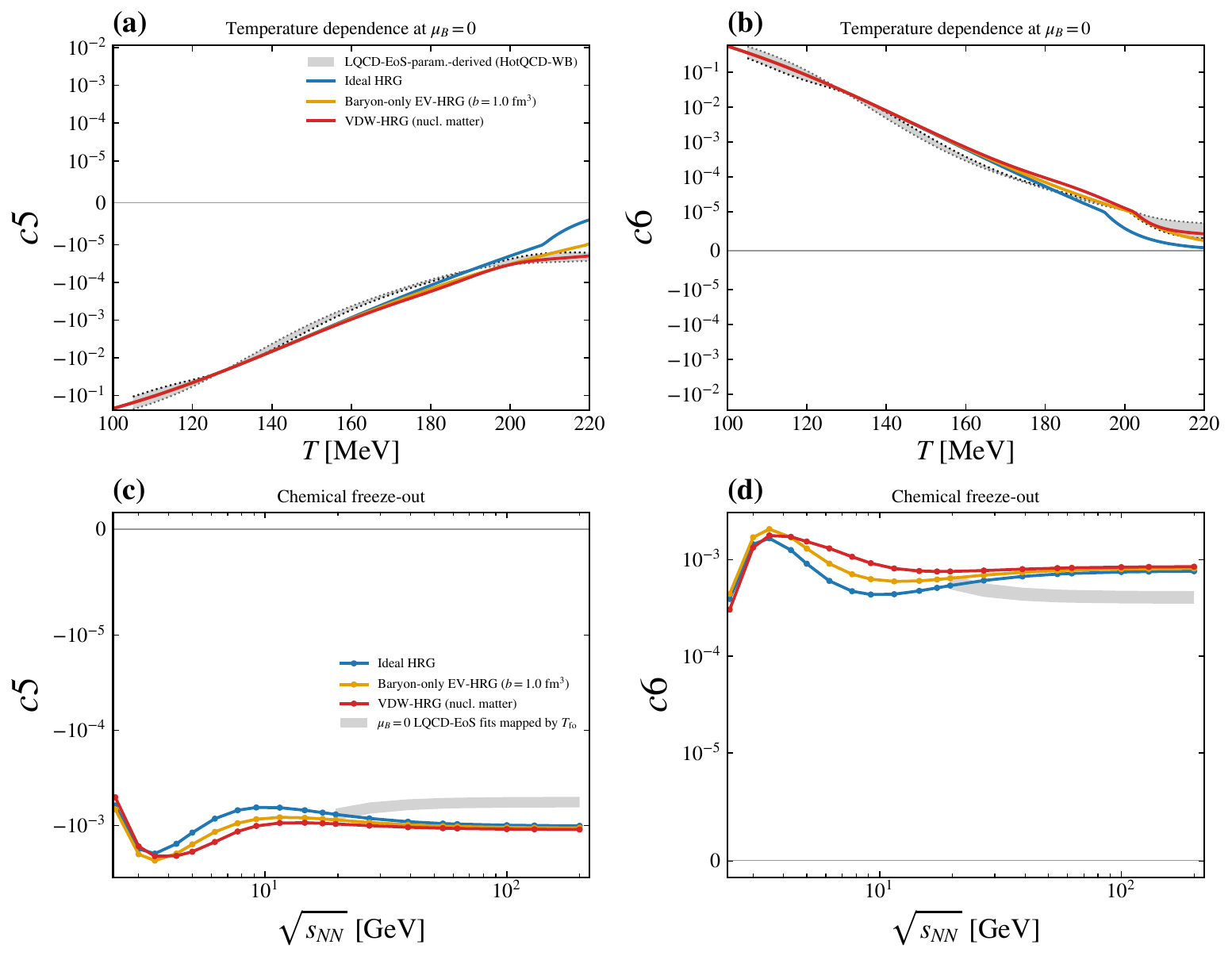}
\caption{Fifth- and sixth-order temperature cumulants. Panels (a) and (b):
$c_5$ and $c_6$ versus $T$ at $\mu_B=0$ for the HRG models and results derived
from the HotQCD and WB EoS parametrizations
\cite{HotQCD:2014kol,Borsanyi:2013bia}. Panels (c) and (d): the HRG results
along chemical freeze-out \cite{Andronic:2017pug}. The grey
high-energy reference maps the $\mu_B=0$ lattice parametrizations through
$T_{\rm fo}$ and is not a finite-density lattice prediction.}
\label{f:c56}
\end{figure*}

Figure~\ref{f:c56} extends the calculation to $c_5$ and $c_6$. At $\mu_B=0$,
shown in Figs.~\ref{f:c56}(a) and (b), all HRG and lattice-parametrization
curves give $c_5<0$ and $c_6>0$ over the displayed range, with the models
staying close on the hadronic side and separating at larger temperature.
Along freeze-out, shown in Figs.~\ref{f:c56}(c) and (d), $c_5$ develops a
broad negative structure and $c_6$ a positive maximum at low collision
energy, with
a stronger model ordering than at lower orders, as expected from the fifth
and sixth derivatives in Eqs.~(\ref{e:c5}) and (\ref{e:c6}).

The grey curves in Fig.~\ref{f:c56} require a stronger qualification than the
$c_2$ points in Fig.~\ref{f:cnT}. They are fifth and sixth derivatives of
analytic functions fitted to lower-order observables, not lattice
measurements of $c_5$ or $c_6$. Differentiation amplifies small differences
between equally acceptable parametrizations, and the spread of the two curves
does not include the covariance of the underlying lattice data. These curves
should be read as guidance for the scale and sign pattern; a direct lattice
determination of the higher temperature derivatives with a controlled
continuum limit would provide the definitive comparison.

\FloatBarrier

\end{document}